**Gate-tunable spin exclusive or operation in a silicon-based spin device at room temperature**

Ryoma Ishihara[1,#], Yuichiro Ando[1,#,†] Soobeom Lee[1], Ryo Ohshima[1], Minori Goto[2], Shinji Miwa[2], Yoshishige Suzuki[2], Hayato Koike[3], and Masashi Shiraishi[1,¶]

[1]Department of Electronic Science and Engineering, Kyoto University, Kyoto, Kyoto, Japan
[2]Graduate School of Engineering Science, Osaka University, Toyonaka, Osaka, Japan
[3]Advanced Products Development Center, TDK Corporation, Ichikawa, Chiba, Japan

**Corresponding authors**

[†]**Yuichiro Ando** Address: A1-226, Kyodai Katsura, Nishigyo-ku, Kyoto, Kyoto, Japan
Tel.: +81-75-383-2356, Fax: +81-75-383-2275
E-mail: ando@kuee.kyoto-u.ac.jp

[¶]**Masashi Shiraishi** Address: A1-222, Kyodai Katsura, Nishigyo-ku, Kyoto, Kyoto, Japan
Tel:+81-75-383-2272, Fax:+81-75-383-2272
E-mail: mshiraishi@kuee.kyoto-u.ac.jp

ABSTRACT

Room temperature operation of a spin exclusive or (XOR) gate was demonstrated in lateral spin valve devices with nondegenerate silicon (Si) channels. The spin XOR gate is a fundamental part of the magnetic logic gate (MLG) that enables reconfigurable and nonvolatile NAND or OR operation in one device. The device for the spin XOR gate consists of three iron (Fe)/cobalt (Co)/magnesium oxide (MgO) electrodes, i.e., two input and one output electrodes. Spins are injected into the Si channel from the input electrodes whose spin angular momentum corresponds to the binary input 1 or 0. The spin drift effect is controlled by a lateral electric field in the Si channel to adjust the spin accumulation voltages under two different parallel configurations, corresponding to (1, 1) and (0, 0), so that they exhibit the same value. As a result, the spin accumulation voltage detected by the output electrode exhibits three different voltages, represented by an XOR gate. The one-dimensional spin drift-diffusion model clearly explains the obtained XOR behavior. Charge current detection of the spin XOR gate is also demonstrated. The detected charge current has a maximum of 0.94 nA, the highest value in spin XOR gates reported thus far. Furthermore, gate voltage modulation of the spin XOR gate is also demonstrated, which enables operation of multiple MLG devices.

## I. INTRODUCTION

Logic gates using electron spins are expected to realize beyond complementary metal-oxide-semiconductor (CMOS) architectures, which exhibit superior switching energy and high logic density compared to the traditional CMOS architecture. Furthermore, they also provide the ability to integrate logic with nonvolatile storage in ferromagnetic memory. Whereas a variety of proposals for logic operation based on spin-dependent phenomena have been presented [1–14], we focus on the semiconductor-based universal magnetologic gate (MLG) proposed by Dery et al. [11] where the operand of logic operation is the magnetization direction. The MLG consists of five elongated ferromagnetic (FM) electrodes with parallel easy magnetization axes lined up on one semiconductor channel (Fig. 1(a)). The two collinear easy axes, $+y$ and $-y$, are defined as the binary states "1" and "0", respectively. The two outmost FM electrodes (FM-A and FM-A') are input terminals, the center electrode (FM-M) is the output terminal, and the other electrodes (FM-B and FM-B') are configuration terminals that define the gate operation (e.g., NAND or OR). The MLG operates as a NAND (OR) gate when the magnetizations of FM-B and FM-B' are both 1 (0). By applying charge currents between FM-A (FM-A') and FM-B (FM-B'), spin accumulation is generated in the semiconductor channel, whose amplitude beneath the FM-M contact is represented by OR(XOR(A, B), XOR(A', B')). The output signal detected by FM-M is a spin-dependent voltage or a spin-dependent charge current. Any binary logic operation can be realized by using a finite number of MLGs. Furthermore, the ability to reconfigurable logic gates at a clock frequency provides flexibility in logic circuit design, which enables a decrease in the number of gates and the time delay. A Boolean expression corresponds to an MLG consisting of two exclusive or (XOR) gates. Therefore, logic operation of one XOR gate using three ferromagnetic electrodes, i.e., FM-A, FM-B and FM-M, is a fundamental technique to realize MLG operation. XOR gate operation has been demonstrated in graphene-based lateral spin devices at room temperature [15]. However, no implementation of XOR operation in a nondegenerate semiconductor, such as silicon (Si), has been demonstrated. Graphene is an atomically thin material, and its physical properties, such as conductivity and carrier types, are strongly affected by adsorbents; i.e., graphene is not tolerant to contamination [16–18]. Meanwhile, Si is quite stable and robust to the surrounding environment in terms of its physical properties. Furthermore, a gate function using a metal-oxide-semiconductor (MOS) structure for modulation of the electron conductivity has been established in Si, which is indispensable for the operation of an MLG with low power consumption. In a practical MLG logic circuit, a clock for simultaneous operation of all MLGs can be constructed in two ways: magnetization rotation of FM-M or gate modulation of the channel conductivity. From the point of view of the energy consumption and magnetic stability of the other ferromagnetic electrodes, the latter method is desired. The energy consumption for constructing one clock in the latter way is estimated to be at least 10 times smaller than that in the former way [19,20]. For Si-based devices, efficient modulation of the conductivity of more than six orders of magnitude has already been established in lateral spin devices, and significant modulations of the electron conductivity and spin accumulation voltage have been demonstrated [21,22]. Modulation of the spin transport length using an electric field, well known as the "spin drift effect", has also been demonstrated both in degenerate and nondegenerate Si spin devices, which is a key technique for XOR operation [22–26]. Furthermore, long-range spin transport due to the low spin scattering probability in the Si channel [27–29] and highly spin polarized spin injection using Fe/MgO epitaxial layers [30] have also been reported, which are general requirements for highly efficient logic operation using a spin current.

In this study, room temperature operation of an XOR gate was demonstrated using Fe/MgO/Si multiterminal lateral spin devices. By adjusting the charge current in the nondegenerate Si channel, which can control the spin drift effect, clear XOR signals are obtained for several devices. In addition, detection of a spin-dependent charge current, i.e., a charge current controlled by the XOR gate, is also demonstrated, which is key for operation of multiple MLGs using a thyristor latch. Furthermore, back-gate modulation of the XOR gate operation is also realized, which is useful for operation of multiple XOR gates.

## II. DEVICE FABRICATION AND NONLOCAL FOUR-TERMINAL MEASUREMENTS

Silicon-on-insulator substrates, consisting of 100-nm-thick Si(100) layer/200-nm-thick buried $SiO_2$ layer/625-μm-thick Si(100) substrate, were employed for fabrication of nondegenerate Si-based multiterminal lateral spin valves (LSVs) for XOR operation. Phosphorus (P) was ion implanted into the Si layer, at a concentration of approximately $1\times10^{18}$ $cm^{-3}$. The conductivity of the Si channel ($\sigma_{Si}$) measured using a conventional four-terminal method was $1.93\times10^{3}$ $(\Omega m)^{-1}$ at 300 K. Prior to deposition of ferromagnetic metal/tunnel barrier layers, a 20-nm-thick highly doped silicon epitaxial layer was grown by magnetron sputtering to suppress the depletion layer thickness. Au (3 nm)/Fe (12.4 nm)/Co (0.6 nm)/MgO (0.8 nm) layers were subsequently deposited on the Si channel by molecular-beam epitaxy. After deposition of the layers, a Si spin channel with three FM contacts (FM-A, FM-B and FM-M) was fabricated by electron-beam lithography and argon-ion ($Ar^+$) milling. The top surface of the Si spin channel was etched to remove 20 nm of the highly doped silicon layer. Finally, two outer nonmagnetic electrodes (NM1 and NM2) were fabricated. Fig. 1(b) shows a schematic of the typical device structure. In the XOR operation, a charge current was applied between FM-A and FM-B, where FM-A (FM-B) was under spin injection (extraction) conditions. The spin accumulation voltage was measured between FM-M and NM2. We fabricated several devices (device A, B and C) with different device geometries. The electrode widths of FM-A ($w_A$), FM-B ($w_B$), and FM-M ($w_M$) were 2, 0.8 and 0.2 μm, respectively. The center-to-center distances between adjacent electrodes ($d_{AB}$ and $d_{BM}$) were $d_{AB}$ = 3.0 μm and $d_{BM}$ = 1.5 μm for device A and B and $d_{AB}$ = 21 μm and $d_{BM}$ = 1.5 μm for device C. All measurements were carried out at 300 K using a direct charge current (dc) technique with a commercial DC source meter and a digital multimeter.

First, we implemented nonlocal four-terminal magnetoresistance (NL-MR) measurements to investigate the spin transport parameters of fabricated devices, in which a magnetic field was applied along the ±$y$ direction to control the magnetization configuration. Fig. 1(c)-(e) show typical NL-MR signals measured at 300 K. The spin injector and detector were (c) FM-A and FM-B, (d) FM-B and FM-M and (e) FM-A and FM-M, respectively. The magnetization directions of each electrode are also displayed in the figures. Clear rectangular signals were detected for all spin injector and detector combinations, indicating that all electrodes have finite spin polarization. The magnetic flux density, $B_y$, for magnetization switching of FM-A, FM-B and FM-M was approximately 7, 16, and 36 mT, respectively. A considerable spin signal was also detected with the FM-A spin injector and FM-M spin detector, even though FM-B is located between the two electrodes, indicating that the spin absorption by FM-B is negligibly small owing to the existence of the MgO tunnel barrier [31–33]. The magnitude of the NL-MR signals, $\Delta V_s$, as a function of the center-to-center distance between the spin injector and detector, $d$, is displayed in Fig. 1(f). $\Delta V_s$ monotonically decreases with increasing $d$. The spin diffusion length $\lambda_s$ was estimated to be 1.54 ± 0.30 μm by using the following fitting function:

$$\Delta V_s = V_0 \exp\left(\frac{-d}{\lambda_s}\right). \quad (1)$$

Successful fitting also indicates negligible spin absorption by FM-B. Hanle effect measurements were also implemented between FM-A and FM-B, in which a magnetic field was applied along the ± $z$ direction. Clear dip and peak features are observed under the parallel and antiparallel configurations, respectively, as shown in Fig. 1(g), indicating successful spin manipulation by the magnetic field. In the analysis, we calculated the difference in the nonlocal voltage between the antiparallel and parallel configurations as shown in Fig. 1(h) and used the following fitting function [34–37]:

$$\frac{V_{\mathrm{NL_AP}}(B_z)-V_{\mathrm{NL_P}}(B_z)}{I} = \frac{P^2\lambda_s}{\sigma_{Si}A}\left(1+\omega^2{\tau_s}^2\right)^{-\frac{1}{4}}\exp\left\{-\frac{d}{\lambda_s}\sqrt{\frac{\sqrt{1+\omega^2{\tau_s}^2}+1}{2}}\right\}\left\{\cos\left(\frac{\arctan(\omega\tau_s)}{2}+\frac{d}{\lambda_s}\sqrt{\frac{\sqrt{1+\omega^2{\tau_s}^2}-1}{2}}\right)\right\}, \quad (2)$$

where $V_{\mathrm{NL_AP}}$ ($V_{\mathrm{NL_AP}}$) is the nonlocal voltage under the antiparallel (parallel) configuration, $B_z$ is the magnetic flux density along the $z$ direction, $P$ is the spin polarization, $A$ is the cross-sectional area of the channel, $\omega = g\mu_B B/\hbar$ is the Larmor frequency, $g$ is the $g$-factor for the electrons ($g$ = 2 in this study), $\mu_B$ is the Bohr magneton, and $\hbar$ is the Dirac constant. From the analysis, the spin diffusion length of the Si channel was estimated to be 1.41± 0.16 μm, which is consistent with the results in Fig. 1(f) and those of previous studies [28].

## III. XOR OPERATION

In the XOR operation, a DC charge current was applied from FM-A to FM-B, and the spin accumulation voltage was measured by FM-M with reference to NM2. The electrochemical potential $\mu$ of up and down spins in the nondegenerate Si channel calculated by the one-dimensional spin drift-diffusion model is shown in Fig. 2(a) and (b), where the potential drop due to the electric field was eliminated for convenience of discussion [26]. In the calculation, the gap distances between FM-A and FM-B ($d_{AB}$) and between FM-B and FM-M ($d_{BM}$) were fixed at 3.0 and 1.5 μm, i.e., FM-A, FM-B and FM-M were located at $x$ = 0, 3.0, and 4.5 μm, respectively. $\lambda_s$, $\sigma_{Si}$ and spin polarization of ferromagnetic electrodes were 1.5 μm, 2000 $(\Omega m)^{-1}$ and 8%, respectively, typical values in our Si spin valves. $\mu$ under the parallel (antiparallel) configuration of FM-A and FM-B is shown in solid (broken) lines. Under the parallel configuration, the directions of the spins injected from FM-A and those left in the Si channel after extraction from FM-B are opposite to each other. The accumulated up (down) spins beneath the FM-A (FM-B) are transported through the Si channel as a spin diffusion current and a spin drift current. As a result, the spin accumulation potential $\Delta\mu_s$, i.e., the difference in $\mu$ between up and down spins, becomes smaller than that under the antiparallel configuration, in which the same orientation spins are accumulated beneath FM-A and FM-B. Here, we focus on $x_{con}$, at which $\Delta\mu_s$ becomes 0 due to the equal numbers of up and down spins transported from FM-A and FM-B, respectively. At $I$ = 50 μA, $x_{con}$ is approximately 2.1 μm, and $\mu$ at FM-M ($x$ = 4.5 μm) under each magnetic configuration ($\mu_{\uparrow\uparrow}$, $\mu_{\downarrow\downarrow}$, $\mu_{\uparrow\downarrow}$ and $\mu_{\downarrow\uparrow}$) represents four different values. In contrast, at $I$ = 200 μA, $x_{con}$ reaches 3.0 μm because of the enhanced spin drift effect, resulting in $\mu_{\uparrow\uparrow} = \mu_{\downarrow\downarrow}$ and three different $\mu$ at $x$ = 4.5 μm. Since the spin accumulation voltage, $V_{XOR}$, detected by FM-M is expressed as $V_{XOR} = P_M \frac{\Delta\mu_s(x=4\mu m)}{e}$, where $P_M$ is the spin polarization of FM-M and $e$ is the elementary charge, it represents ternary values at $I$ = 200 μA. Fig. 2(c) shows the current dependence of $\mu$ at $x$ = 4.5 μm. $\mu_{\uparrow\downarrow}$ ($\mu_{\downarrow\uparrow}$) monotonically increases (decreases) with increasing charge current. In contrast, $\mu_{\downarrow\downarrow}$ ($\mu_{\uparrow\uparrow}$) first increases (decreases), then decreases (increases) to 0 meV at a specific current $I_0$ and finally changes its polarity and decreases (increases). Fig. 2(d)-(f) show the expected spin signals at $I < I_0$, $I = I_0$, and $I > I_0$ under the application of a magnetic field along the $y$ direction. Here, we suppose that the magnetization switching field of FM-M is designed to be higher than those of FM-A and FM-B and fixed along the $-y$ direction during the measurements. In the following, we label the spin accumulation voltage under each magnetic configuration by $V_{\uparrow\uparrow}$, $V_{\downarrow\downarrow}$, $V_{\downarrow\uparrow}$ and $V_{\uparrow\downarrow}$, where the left (right) suffix is the magnetization direction of FM-A (FM-B). Since the polarity of $V_{\uparrow\uparrow} - V_{\downarrow\downarrow}$ depends on $I$, the shape of the hysteresis is drastically changed by changing $I$. These features of the hysteresis curves are expected in the spin XOR operation. Although the conventional nonlocal magnetoresistance can also be recognized as the XOR operation because the spin accumulation voltage is different between the parallel and antiparallel configurations, this feature is not applicable for the NAND or OR operation in an MLG.

The magnetic field dependence of the spin accumulation voltage, $V_{\mathrm{XOR}}$, measured between FM-M and NM2 at 300 K is shown in Fig. 3(a)-(c). A charge current was applied from FM-B to FM-A. The magnetization of FM-M was fixed along the -$y$ direction. The applied $B_y$ was in the range between -30 and 30 mT, which is sufficiently small for magnetization switching of FE-M (36 mT), as confirmed in Fig. 1(c)-(e). For device A, $V_{\uparrow\uparrow}$, is less than $V_{\downarrow\downarrow}$ at $I$ = 0.1 mA, corresponding to Fig. 2(d), indicating an insufficient spin drift effect. At $I$ = 0.3 mA≈ $I_0$, $V_{\uparrow\uparrow} = V_{\downarrow\downarrow}$ is realized, resulting in the $V_{\mathrm{XOR}}$-$B_{\mathrm{y}}$ curve with ternary values, a successful demonstration of the XOR operation. At a charge current higher than $I_0$, the $V_{\mathrm{XOR}}$-$B_{\mathrm{y}}$ curve shows four different values with $V_{\downarrow\downarrow} < V_{\uparrow\uparrow}$, corresponding to Fig. 2(f). Although similar signals were obtained at $I$ = 0.3 and 1.2 mA for device B, the signal-to-noise (S/N) ratio became worse below 0.2 mA, and no signals were detected at 0.1 mA because of the small charge current. Despite having the same device geometry, devices A and B exhibit several differences. First, the signal amplitude (the difference between $V_{\downarrow\uparrow}$ and $V_{\uparrow\downarrow}$) of device A is larger than that of device B for all $I$ conditions, and a clear signal was detected even at $I$ = 0.1 mA for device A. Second, $V_{\downarrow\downarrow}$ is slightly larger than $V_{\uparrow\uparrow}$ for device B at $I$ = 0.3 mA, indicating that the $I_0$ condition is slightly shifted. Such a difference might be due to the slight difference in the spin polarization of the ferromagnetic electrodes and/or $\sigma_{\mathrm{Si}}$ of the Si channel. When the spin polarization of FM-M is decreased, the signal magnitude also decreases for all $I$ regions and finally drops to below the detection limit. In contrast, when the spin polarization of FM-A or FM-B or the conductivity in the Si channel are modulated, the $I_0$ condition is changed. When the spin polarization of FM-B is changed from 8 % to 10 % in the situation of Fig. 2, the $I_0$ condition is estimated to change from 0.20 to 0.23 mA. Therefore, precise control of the spin polarization of the ferromagnetic electrodes and $\sigma_{\mathrm{Si}}$ of the channel are strongly desired for reliable operation of multiple MLGs. To solve such demanding requirements, a new way to adjust the $I_0$ condition even for devices with scattered spin polarizations and/or $\sigma_{\mathrm{Si}}$ will be discussed in section V.

We also demonstrated XOR operation with long $d_{\mathrm{AB}}$ (Fig. 3(c)). A clear XOR-operated signal was obtained at $I = I_0$ = 0.6 mA. Since $d_{\mathrm{AB}}$ is 21 μm for device C, a large $I_0$ (0.6 mA) was obtained due to a further spin drift effect that shifted the $x_{\mathrm{con}}$ point to $x$ = 21 μm. The long-distance XOR operation enables the addition of several ferromagnetic electrodes between FM-A and FM-B to realize a high degree of design freedom. Because of the negligible spin absorption by the ferromagnetic electrode, we can freely add additional spin injectors. Spin logic gates other than the MLG, such as a majority circuit, can be realized using multiple ferromagnetic electrodes. The charge current dependences of $V_{\uparrow\uparrow}$, $V_{\downarrow\downarrow}$, $V_{\downarrow\uparrow}$ and $V_{\uparrow\downarrow}$ are summarized in Fig. 4 for devices B and C, where $(V_{\downarrow\uparrow} + V_{\uparrow\downarrow})/2$ was subtracted as a background voltage. The behaviors qualitatively correspond to the theoretical behavior shown in Fig. 2(c).

## IV. CHARGE CURRENT DETECTION OF THE XOR GATE

In the MLG operation, a spin-dependent charge current is expected as an output signal to operate the next MLG via a thyristor. Therefore, detection of the XOR-operated charge current, $I_{XOR}$, was also carried out. The current-voltage configuration is shown in Fig. 5(a). Instead of the digital multimeter, a resistor, $R_{XOR}$, was inserted between FM-M and NM2. The resistance of $R_{XOR}$ was (b) 100 Ω, (c)1 kΩ and (d) 100 kΩ. The resistance of $R_{XOR}$ is much smaller than that of the internal resistor of the digital multimeter (>10 GΩ). $I_{XOR}$ was calculated from the voltage drop at $R_{XOR}$. The results for device B are shown in Fig. 5(b)-(d), where $I$ was 0.3 mA. A clear XOR operation is also demonstrated even for the current detection conditions. The magnitude of the output signals, $\Delta I_{XOR}$, i.e., the difference in $I_{XOR}$ between two different antiparallel conditions, is 0.94 nA at $R_{XOR} = 100\ \Omega$ , the highest value reported thus far. Except for the magnitude of the signals, no significant change was found over a wide range of $R_{XOR}$. The magnitude of the output signals as a function of $R_{XOR}$ is summarized in Fig. 5(e). We used the following fitting function:

$$\Delta I_{XOR} = \frac{V_{\uparrow\downarrow} - V_{\downarrow\uparrow}}{R_{XOR} + R_{device}}, \quad (3)$$

where $R_{device}$ is the two terminal resistance between FM-M and NM2. The fitting curve nicely reproduces the experimental data. $V_{\downarrow\uparrow} - V_{\uparrow\downarrow}$ and $R_{device}$ are estimated to be 14 ± 5 μV and 16 ± 6 kΩ. The value of $R_{device}$ is consistent with the resistance at zero bias in the previous studies [26]. Because of the considerable parasitic resistance, the maximum $\Delta I_{XOR}$ is expected to be ca. 1 nA, less than the shot noise level in the operation at 1 GHz [11]. However, a 10~100-fold larger $\Delta I_{XOR}$ is possible by reducing the parasitic resistance and the gap distances $d_{AB}$ and $d_{BM}$.

## V. OPERATION OF A GATE-TUNABLE XOR LOGIC GATE

In section III, we reported scattered $I_0$ probably due to the scattered spin polarization and/or $\sigma_{Si}$ despite the same device geometries. In this section, we demonstrate gate modulation of the XOR operation to control the $I_0$ condition. A back gate, $V_G$, was applied to control $\sigma_{Si}$ and the spin drift effect. The $V_G$ dependence of $\sigma_{Si}$ is displayed in Fig. 6(a). $\sigma_{Si}$ was modulated by more than ten times by application of ±30 V (electric field, $E_G$ = ±150 MV/m). $V_G$-dependent $V_{XOR}$-$B_y$ curves at $I$ = 0.3 mA for device B are shown in Fig. 6(b)-(d). A small linear background, expressed as $V_{XOR} = A \times B_y$, was subtracted from the raw data to clarify the difference between $V_{\downarrow\downarrow}$ and $V_{\uparrow\uparrow}$. Note that such a treatment does not affect subsequent discussions because the same linear background was subtracted from the data of the up and down sweeps. At $V_G = 0$ V, although a clear signal was detected, $V_{\downarrow\downarrow}$ is slightly larger than $V_{\uparrow\uparrow}$ due to a slight deviation from the $I_0$ condition. At $V_G = -15$ V, the voltage difference between $V_{\downarrow\downarrow}$ and $V_{\uparrow\uparrow}$ was successfully decreased. In contrast, upon application of $V_G = +15$ V, the difference became pronounced. The voltage difference between $V_{\downarrow\downarrow}$ and $V_{\uparrow\uparrow}$ as a function of $V_G$ is shown in Fig. 6(e), where $V_{\downarrow\downarrow}$ and $V_{\uparrow\uparrow}$ were calculated by averaging the data between -5 mT and +5 mT. The voltage difference was systematically modulated by $V_G$, indicating successful demonstration of gate modulation of the XOR gate.

Hereafter, we discuss the advantages of the gate modulation of the XOR gate. In practical spin current logic gates such as MLGs, many devices should be combined. For reliable operation, $I_0$ should be designed with the same value. However, $I_0$ is scattered because of the immaturity of device fabrication technology, as discussed in section III. Even for the mature Si CMOS technology, this problem becomes pronounced because of the dispersion of the number of dopant atoms, which impedes further progress in Moore's law. In contrast, in our spin current device, the threshold condition can be controlled after fabrication by using a gate function, which is a great advantage over the conventional CMOS device. If a floating gate is fabricated on the Si channel, then such an adjustment is easily realized with nonvolatility. Such a threshold modulation can be useful similar to the body effect in a conventional MOS field-effect transistor. In Fig. 6, gate modulation of the XOR operation was demonstrated under constant current application. Under the constant voltage condition, both the $I_0$ condition and the magnitude of the spin signal are expected to change. Such a modulation is also useful for weighting algorithms controlled by floating gates.

## VI. CONCLUSION

In conclusion, we have demonstrated room temperature XOR operation in Fe/MgO/Si lateral spin devices. By adjusting the in-plane electric field in the Si channel, clear XOR-operated hysteresis signals have been detected for several devices with different channel lengths. XOR operation over a long channel distance of 21 μm enables a high degree of design freedom for multiterminal ferromagnetic electrodes. Charge current detection of the XOR operation has also been demonstrated. Furthermore, the charge current condition for the XOR operation has been modulated by a gate function, which is a great advantage for multiple MLG operation. Whereas an external magnetic field has been employed to control the magnetic configuration in this study, individual control of each magnetization in multiple ferromagnetic electrodes becomes difficult, such as in an MLG. For the practical use of MLGs, employment of spin-orbit torque or voltage-induced magnetization switching is desired [38–41]. Demonstration of a spin logic gate by using the magnetic proximity effect instead of the conventional electrical spin injection through ferromagnetic metals is also worthy of study. Several studies have reported that the sign and magnitude of the spin polarization in the magnetically proximitized layer can be controlled by the gate function [42–45]. Such a sign reversal of the spin polarization can be useful for ultrafast switching of the logic configuration in an MLG at a clock frequency.

ACKNOWLEDGMENTS

This work was supported by JSPS (KAKENHI No. 16H06330 and No. 19H02197).

AUTHOR CONTRIBUTIONS

#These authors contributed equally to this work

FIGURE CAPTIONS

FIG. 1. (a) Schematic illustration of the semiconductor-based MLG device proposed by Dery et al. [11]. (b) Schematic illustration of the silicon-based multiterminal LSVs for XOR operation. (c-e) Nonlocal four-terminal magnetoresistance measured at 300 K. The spin injector and detector are (c) FM-A and FM-B, (d) FM-B and FM-M, and (e) FM-A and FM-M, respectively. (f) Gap distance, $d$, dependence of the magnitude of NL-MR signals. The dots are experimental data, and the red line is a fitting result obtained using Eq. (1). (g) Hanle effect signals under parallel (red) and antiparallel (bule) configurations, where the linear background was subtracted. (f) Difference in the Hanle signal between the antiparallel and parallel configurations. The dots and solid line are the experimental data and the fitting curve obtained using Eq. (2).

FIG. 2. Electrochemical potentials $\mu$ of up and down spins in the Si channel calculated by the one-dimensional spin drift-diffusion model. The applied charge current from FM-B to FM-A was (a) 50 and (b) 200 μA. The potential drop due to charge current flow was subtracted. The center-to-center distances between FM-A and FM-B ($d_{AB}$) and between FM-B and FM-M ($d_{BM}$) were 3.0 and 1.5 μm, respectively. The conductivity of the Si was 2000 $(\Omega m)^{-1}$, and the spin diffusion length in the Si channel was 1.5 μm. (c) Charge current dependence of $\mu$ under various magnetic configurations of FM-A and FM-B, i.e., two different parallel configurations (↑↑ or ↓↓) and antiparallel configurations (↑↓ or ↓↑). (d-f) Expected $V_{XOR}$-$B_y$ signal shapes detected by FM-M at (d) $I < I_0$, (e) $I = I_0$ and (f) $I > I_0$.

FIG. 3. $V_{XOR}$-$B_y$ curves at various charge currents for (a) device A, (b) device B and (c) device C, measured at room temperature. The distance between FM-A and FM-B was (a) 3.0, (b) 3.0 and (c) 21 μm. The magnetization of FM-M was fixed along the $-y$ direction. The external magnetic flux density, $B_y$ was swept between -30 to +30 mT, which is sufficiently small for magnetization switching of FM-M.

FIG. 4. Charge current dependence of $V_{\uparrow\uparrow}$, $V_{\downarrow\downarrow}$, $V_{\downarrow\uparrow}$ and $V_{\uparrow\downarrow}$ for (a) device B and (b) device C. ($V_{\downarrow\uparrow}$ + $V_{\uparrow\downarrow}$)/2 was subtracted as a background voltage.

FIG. 5. (a) Schematic illustration of the current-voltage configuration for detection of the XOR-operated charge current. (b - d) XOR-operated charge current, $I_{XOR}$, as a function of $B_y$ measured at 300 K. The resistance of $R_{XOR}$ was (b) 100 Ω, (c) 1 kΩ and (d) 100 kΩ. The applied charge current was 0.3 mA. (f) $\Delta I_{XOR}$ as a function of $R_{XOR}$. The dots and solid line are the experimental data and the fitting curve obtained using Eq. (3).

FIG. 6. (a) Back-gate voltage, $V_G$, dependence of the conductivity of the Si channel measured at 300 K. (b-d) $V_{XOR}$-$B_y$ curves at $V_G$ = (b) -15, (c) 0 and (d) +15 V. A linear background was subtracted for comparison of the voltages between $V_{\uparrow\uparrow}$ and $V_{\downarrow\downarrow}$. (e) Difference between $V_{\downarrow\downarrow}$ and $V_{\uparrow\uparrow}$, as a function of $V_G$.

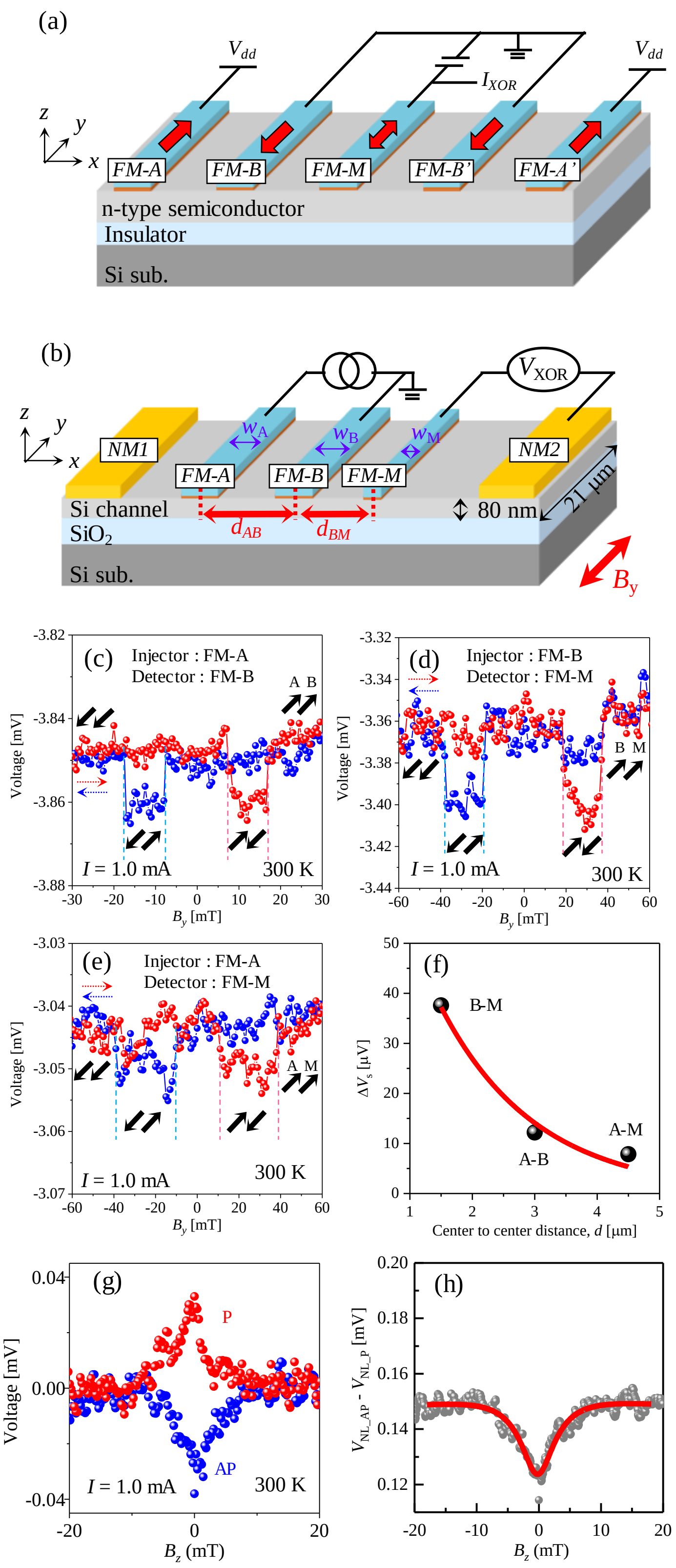

Figure 1 Ishihara et al

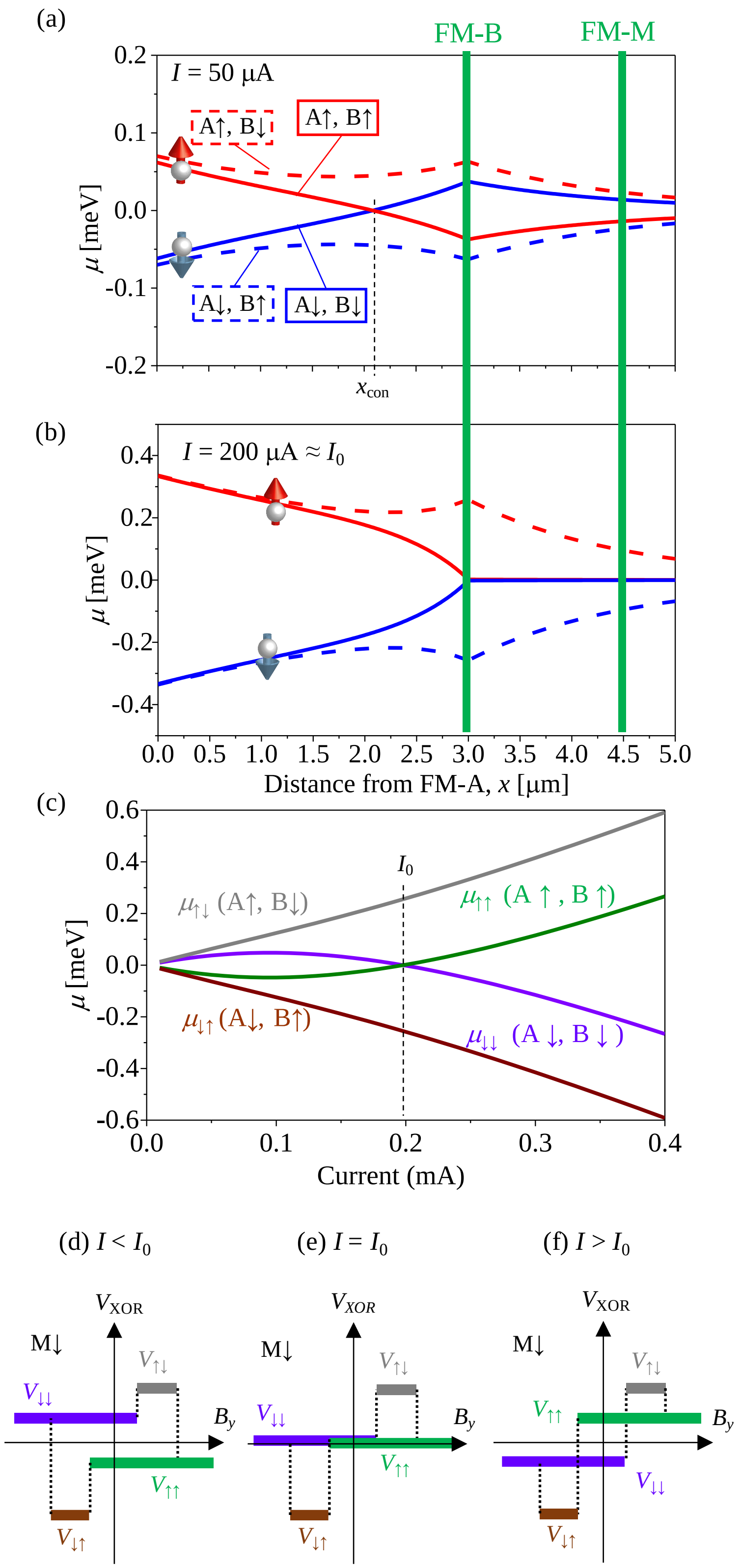

Figure 2 Ishihara et al

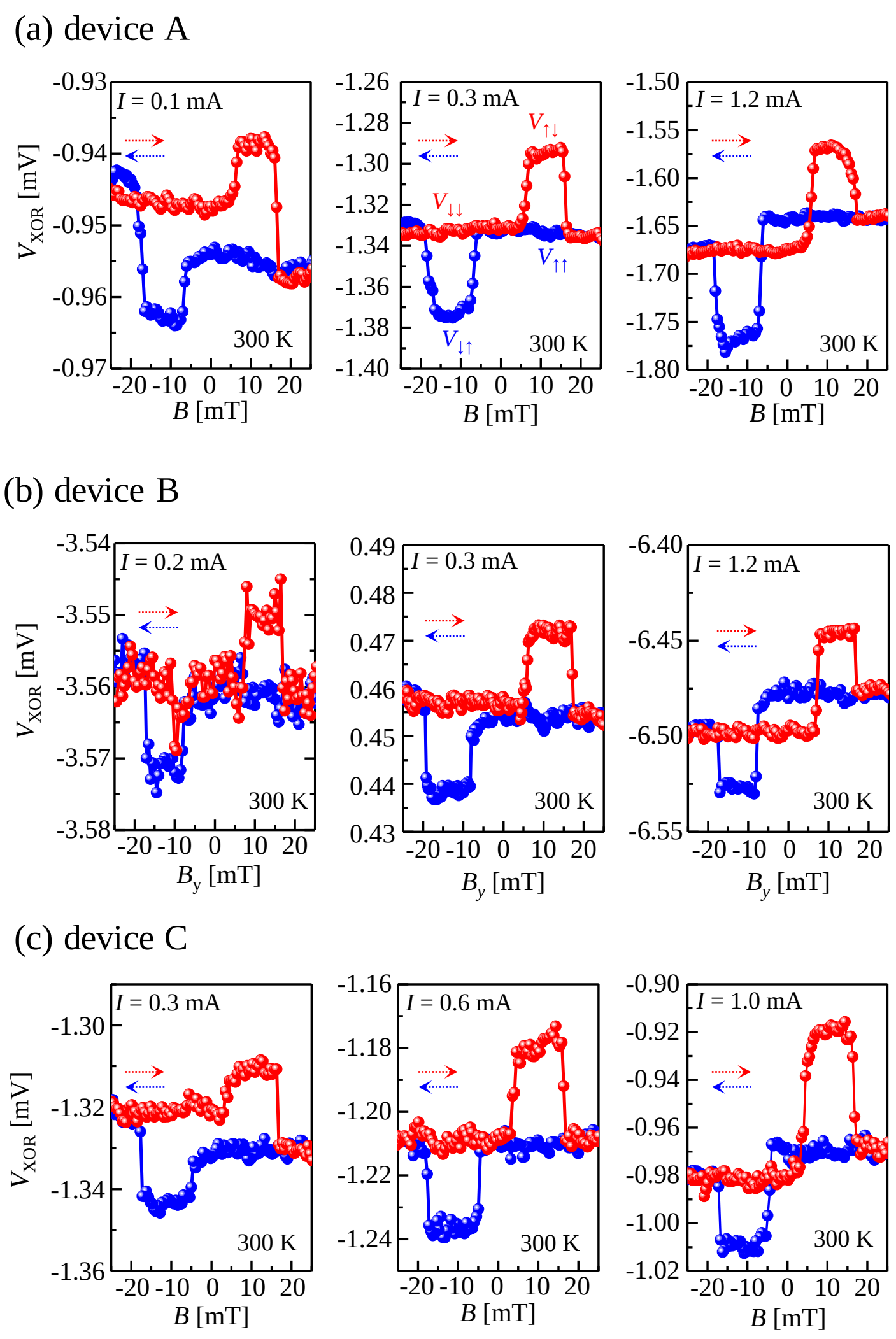

Figure 3 Ishihara et al

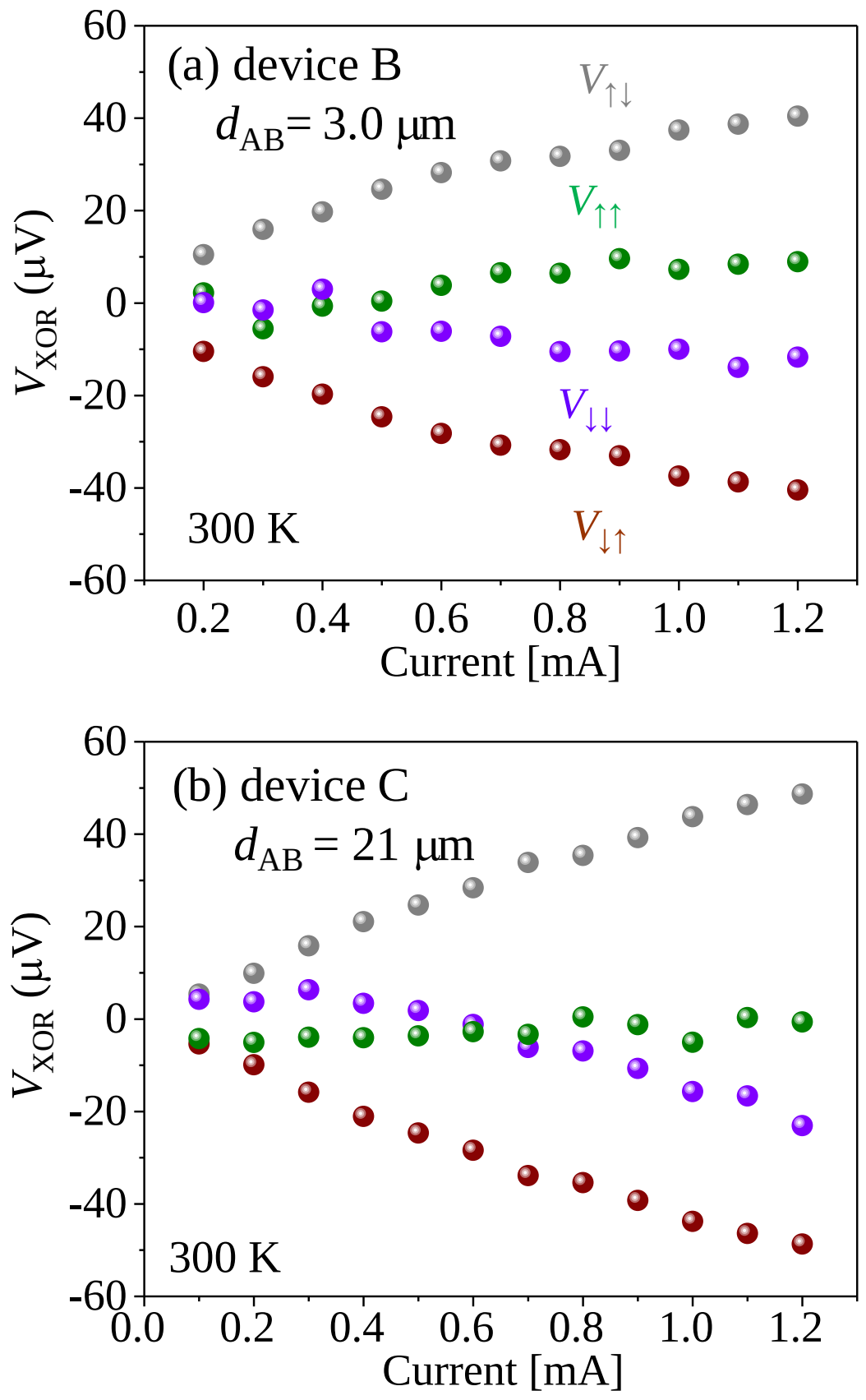

Figure 4 Ishihara et al

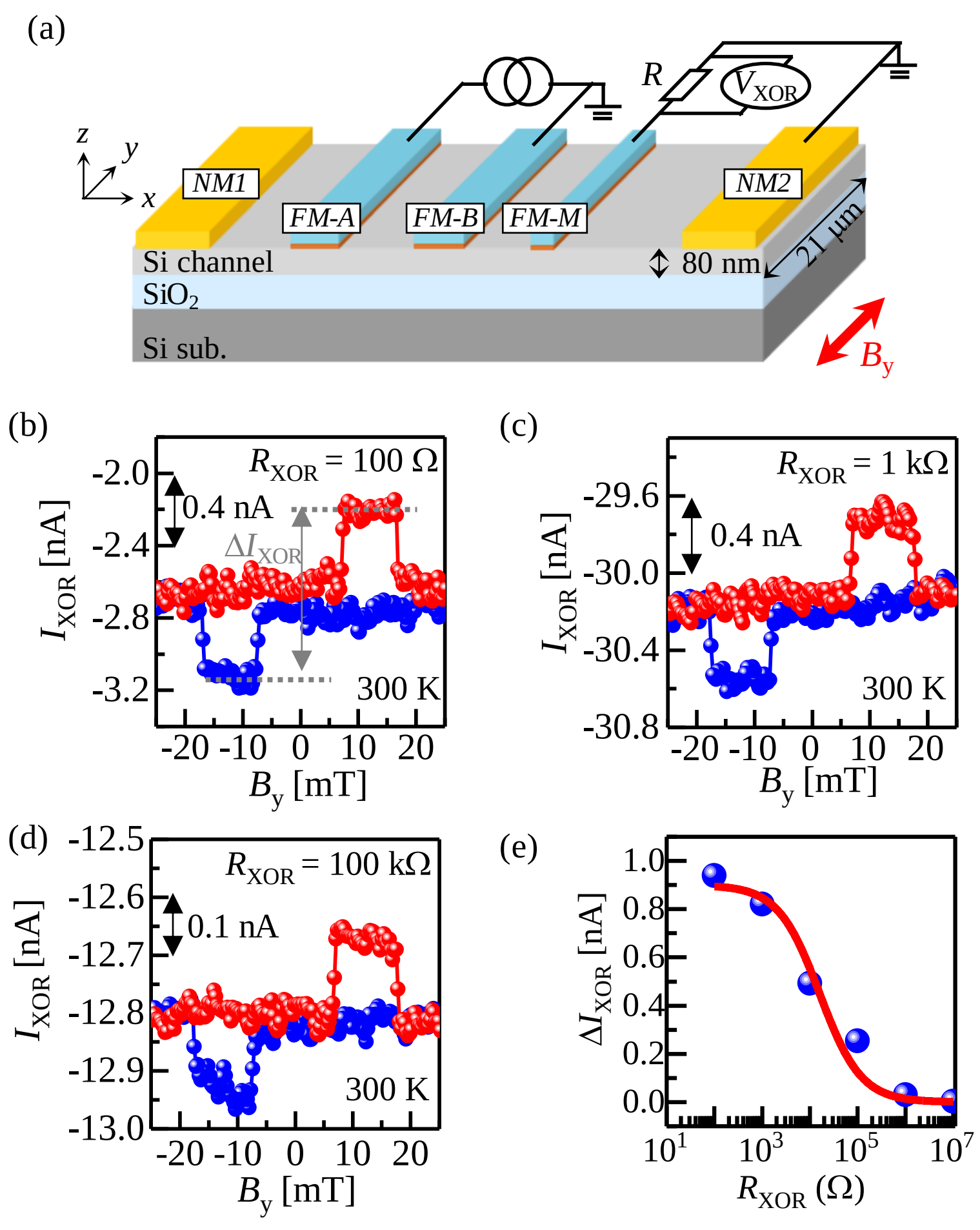

Figure 5 Ishihara et al

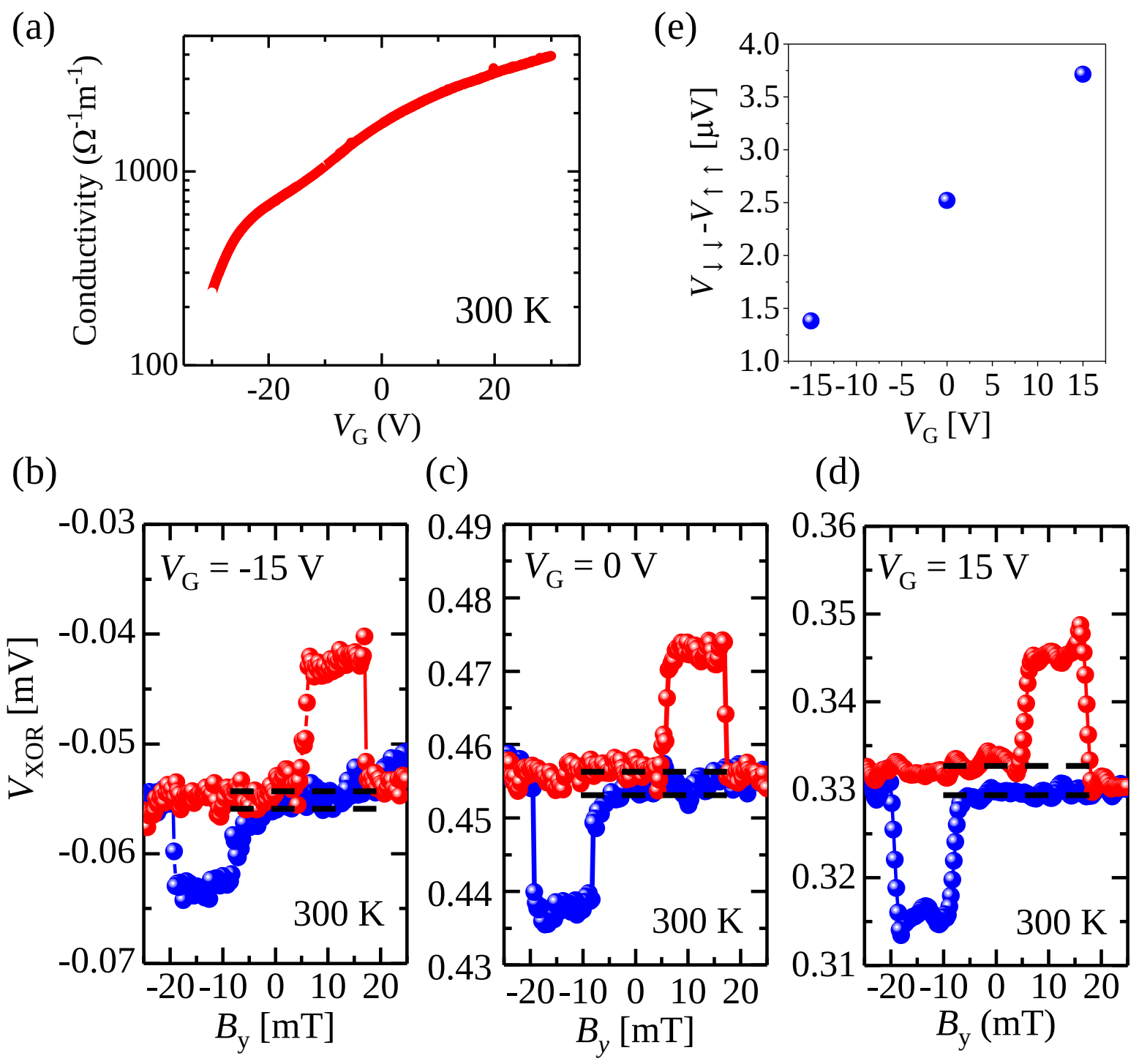

Figure 6 Ishihara et al